# Training Under Attentional Competition Produces Persistent Biases in Visual Appearance

Thitaporn Chaisilprungraung[1], Prapasiri Sawetsuttipan[1,2,3,4], John T Serences[5,*], and Sirawaj Itthipuripat[1,2,6,7,*, ✉]

[1]Neuroscience Center for Research and Innovation (Nx), Learning Institute King Mongkut's University of Technology Thonburi, Bangkok, 10140, Thailand
[2]Big Data Center (Bx), King Mongkut's University of Technology Thonburi, Bangkok, 10140, Thailand
[3]Computer Engineering Department, Faculty of Engineering, King Mongkut's University of Technology Thonburi Bangkok, 10140, Thailand
[4]Division of Information Technology, Office of the President, Mahidol University, Nakhon Pathom 73170, Thailand
[5]Neurosciences Graduate Program, Department of Psychology and Kavli Foundation for the Brain and Mind, University of California, San Diego, La Jolla, California 92093, USA
[6]Rotman Research Institute at Baycrest Academy for Research and Education, Toronto, ON M6A 2E1, Canada
[7]Brainspoke Lab, Alphie, 159836, Singapore

*Equal contributions as senior authors

✉Correspondence to
Sirawaj Itthipuripat (itthipuripat.sirawaj@gmail.com)
Neuroscience Center for Research and Innovation, Learning Institute, King Mongkut's University of Technology Thonburi, Bangkok, Thailand, 10140

**Abstract**

Selective attention can momentarily alter visual appearance, but can such effects be learned? We tested whether training attention under sensory competition produces lasting changes in perceived contrast. Across seven days, participants trained on an orientation task with a fixed target location, with or without a salient distractor. Before and after training, we measured the point of subjective equality (PSE). Training under competition produced a reliable push-pull shift: stimuli at the trained location appeared higher in contrast, whereas stimuli at the untrained location appeared lower. Conversely, training without distractors improved performance but did not alter appearance. Crucially, these opponent shifts were robust to task variations, persisting even in equality judgments designed to minimize response bias. Furthermore, the effect generalized to stimuli with novel orientation and contrast levels. These findings demonstrate that resolving sensory competition does not merely improve discrimination, but durably recalibrates the subjective appearance of the visual world.

## Introduction

Our visual experience feels stable and faithful to the physical world. Yet decades of research show that what we perceive depends not only on sensory input but also on how attention is allocated. When attention is directed to a location, objects there can appear higher in contrast, sharper, or more salient than physically identical objects elsewhere (Carrasco et al., 2004; Anton-Erxleben & Carrasco, 2013; Liu et al., 2009; Fortenbaugh et al., 2011; Gobell & Carrasco, 2005). These findings demonstrate that attention can modulate subjective appearance, that is, how stimuli look rather than merely how efficiently they are detected or discriminated (Anton-Erxleben et al., 2010; Itthipuripat et al., 2023; Itthipuripat, Chang, et al., 2019). However, nearly all such effects have been demonstrated using brief, transient cues and measured on a trial-by-trial basis. Whether repeated experience allocating attention can induce enduring changes in subjective perception remains largely unknown.

A defining feature of attention in natural vision is that it operates under competition. Relevant information must be selected while irrelevant inputs are ignored. Influential theoretical frameworks describe this process in terms of spatial priority maps, in which topographic representations assign greater weight to some locations than others, thereby biasing visual processing toward selected inputs (Desimone & Duncan, 1995; Bisley & Goldberg, 2010; Gottlieb & Balan, 2010; Jerde & Curtis, 2013; Sprague & Serences, 2013; Sprague et al., 2015). In many accounts, this prioritization is constrained by normalization or resource limits, such that enhancing processing at one location is accompanied by reduced influence from competing locations (Reynolds et al., 1999; Martínez-Trujillo & Treue, 2002; McAdams & Maunsell, 1999; Moran & Desimone, 1985). At the neural level, attentional selection alters response gain, tuning selectivity, population-level correlations, and even receptive-field positions. Together, these changes produce location-biased representations that support improved behavioral performance (Connor et al., 1997; Cohen & Maunsell, 2009; Itthipuripat, Ester, et al., 2014; Itthipuripat, Garcia, et al., 2014; Mitchell et al., 2009; Itthipuripat et al., 2017; Itthipuripat & Serences, 2016; Sookprao et al., 2024). If such competitive weighting is engaged repeatedly through experience, it could plausibly leave a lasting imprint on perception itself. This imprint could include not only enhanced influence from consistently selected locations but also diminished perceptual influence from locations that are consistently ignored.

Evidence from the perceptual learning literature supports the possibility that repeated practice can induce durable, location-specific changes in visual processing (Sagi, 2011; Watanabe & Sasaki, 2015). Training on simple discrimination tasks often improves performance most strongly at trained retinal locations and stimulus features, with limited transfer to untrained contexts (Bao et al., 2010; Jehee et al., 2012; Byers & Serences, 2014; Itthipuripat et al., 2017). These improvements are commonly interpreted as reflecting experience-dependent reweighting of sensory signals at the level of decision or readout processes (Dosher & Lu, 1999, 2017). Classic findings show stronger transfer when training and test share representational and decisional components, and weaker transfer when overlap is minimal. However, methods such as double training can reveal generalization that would otherwise remain hidden (Xiao et al., 2008; Wang et al., 2012, 2014). Critically, perceptual learning studies have focused almost exclusively on accuracy and sensitivity, namely how well stimuli are discriminated. This focus leaves unresolved whether learning can also alter subjective appearance, meaning how stimuli are experienced perceptually rather than how effectively they support task performance.

Research on attention-induced changes in appearance provides a complementary perspective. When attention is transiently cued to a location, stimuli there are perceived as higher in contrast or spatial frequency than unattended stimuli, even when physical properties are matched (Carrasco et al., 2004; Anton-Erxleben et al., 2010; Gobell & Carrasco, 2005). Although some of these effects can be influenced by decisional bias, converging evidence from equality judgments and computational modeling indicates a genuine perceptual component (Anton-Erxleben et al., 2010; Itthipuripat et al., 2023). Importantly, however, these appearance changes are short-lived and dissipate once attentional cues are removed. Thus, while transient attention can momentarily bias appearance, it remains unclear whether learning to deploy attention, particularly under sustained competition, can produce enduring changes in subjective perception that persist beyond the training context.

Here, we bridge these literatures by asking a simple but untested question: Can repeatedly resolving attentional competition durably recalibrate subjective appearance? We hypothesized that training attention under competition would bias spatial priority in a push–pull manner, yielding enduring changes in the relative perceptual influence of information across space. Crucially, we do not propose that training alters the physical encoding of contrast or induces permanent changes in early sensory representations. Rather, we test whether repeated competitive selection biases how sensory information from different locations contributes to appearance judgments. Under this account, stimuli at consistently selected locations should come to appear higher in contrast, whereas stimuli at consistently ignored locations should appear lower, even when physical stimulus properties and task demands are equated.

To test this hypothesis, participants trained for five days on an orientation discrimination task in which the target consistently appeared in one visual hemifield. One group trained under sensory competition, with a salient distractor presented in the opposite hemifield on half of the trials, whereas a comparison group trained without distractors. Before and after training, participants completed independent contrast-judgment tasks that measured the point of subjective equality (PSE) between stimuli presented at trained and untrained locations. These test tasks were designed to be independent of the training contingencies and to minimize spatial priors or response habits, allowing us to assess whether learning altered perceived contrast rather than merely biasing decisions.

Across two experiments, we show that training attention under competition produces an enduring push–pull bias in subjective appearance. Only participants who trained with distractors exhibited opponent shifts in perceived contrast. Stimuli at the trained location were subsequently perceived as higher in contrast, whereas stimuli at the untrained location were perceived as lower. These effects persisted beyond the training phase, replicated across judgment formats designed to reduce decisional bias, and generalized partially to novel stimulus contexts. In contrast, training without distractors improved discrimination performance but did not alter appearance. Together, these findings indicate that subjective visual experience is not fixed but can be durably recalibrated by the history of resolving sensory competition.

## Experiment 1

**Materials and Method**

**Participants**

We recruited twenty undergraduate students from the University of California, San Diego (14 female; age = 18-25 years old; one left-handed) to participate in Experiment 1. All participants reported normal or corrected-to-normal vision and provided written informed consent. All procedures were approved by the UCSD Institutional Review Board and conducted in accordance with the ethical standards of the Declaration of Helsinki. Participants received compensation of $10 USD per hour for their participation.

We set a target sample of $N = 20$ for a 2 (Training Condition: distractor-present vs. distractor-absent; between subjects) × 2 (Day: pre vs. post; within subjects) × 2 (Test Location: trained vs. untrained; within subjects) design. A power analysis for the critical Training × Day × Location interaction ($\alpha = .05$, two-tailed; desired power = .80), assuming a medium effect ($f = .25$), a conservative within-subject correlation of $r = .50$, and $\varepsilon = 1$, indicated a total sample of approximately 20–24 participants. For planned pre-to-post comparisons within each training group at the trained location, 10 participants per group provides ~.80 power to detect $d_z \approx 0.75$–$0.80$ (medium-to-large) shifts in the point of subjective equality (PSE)—effect sizes commonly observed in tightly controlled appearance paradigms. This target also aligns with typical sample sizes in comparable psychophysical studies while preserving power for the between-group manipulation (Anton-Erxleben et al., 2010; Carrasco et al., 2004; Itthipuripat, Vo, et al., 2019).

Before data collection, participants were randomly assigned to one of four training routines (5 participants per routine) defined by Training Condition (distractor-present vs. distractor-absent) and Trained Side (left- vs. right-attended; see Fig. 1b).

**Procedure**

Participants completed seven sessions across seven days: a pre-training session (Day 1), five training sessions (Days 2–6), and a post-training session (Day 7). On Days 1 and 7, each participant performed three psychophysical tasks in a counterbalanced order that was held constant within participant across sessions: (i) an orientation discrimination task (ODT; Experiment 1a), (ii) a contrast judgment task (CJT) on oriented grating stimuli; Experiment 1b), and (iii) a CJT on the checkerboard stimuli (Experiment 1c). During training (Days 2–6), participants performed a modified ODT with the target fixed to a single hemifield (left or right) and, depending on group, with or without a high-contrast salient distractor.

**Experiment 1a (ODT).** On each trial, a low-contrast grating (6% Michelson contrast; spatial frequency = 3 c/°, SD of the Gaussian envelope = 2.18°, stimulus radius = 6.53°) appeared at eccentricity =13.74° on either the left or right of fixation. Participants indicated whether the grating was tilted clockwise or counterclockwise from vertical using the index/middle fingers of the dominant hand (Fig. 1a). Two tilt offsets were used—2° (hard) and 4° (easy). On 50% of trials, a high-contrast distractor (90% Michelson) appeared in the opposite hemifield. During the test sessions (Days 1 and 7), target side alternated by half-block (left in the first half, right in the second, or vice versa), and block order was counterbalanced across participants. (During training, the target remained on the assigned side throughout; see Fig.1)

**Experiment 1b (CJT with grating stimuli).** Two gratings (spatial frequency = 3 c/°, SD of the Gaussian envelope = 2.18°, stimulus radius = 6.53°, eccentricity =13.74°) were presented simultaneously, one in each hemifield. One was the standard (fixed at 6% contrast), and the other was the test (contrast randomly drawn from 0.5%, 1.5%, 3%, 4%, 5%, 6%, 8%, 12%, 16%, 24%). Participants made two responses: they first indicated which stimulus appeared higher in contrast using the hand on the same side as that stimulus, and then reported its tilt direction (clockwise/counterclockwise) using the index or middle finger (Fig. 1a, bottom left). Tilt offsets again were 2° and 4°. As in Experiment 1a, the test-stimulus side alternated half-way through each block, with block order counterbalanced.

**Experiment 1c (CJT with checkerboard stimuli).** The display matched Experiment 1b except that the stimuli were checkerboard patterns (spatial frequency = 3 c/°, SD of the Gaussian envelope = 2.18°, stimulus radius = 6.53°, eccentricity =13.74°; Fig. 1a, bottom right). Participants indicated which pattern had higher contrast or whether the two were of equal contrast, using the index, middle, or ring finger of the dominant hand. The option to select equal contrast was included to mitigate decisional bias where subjects feel compelled to select the attended stimulus as higher contrast if forced to choose (Anton-Erxleben et al., 2010). All remaining parameters mirrored Experiment 1b.

**Training (Days 2–6).** Participants completed a modified ODT in which the target grating always appeared on the assigned side (left or right) for a given subject. In the distractor-present group, a 90%-contrast distractor appeared in the opposite hemifield on 50% of trials (randomly interleaved). In the distractor-absent group, no distractors were shown. Tilt offsets were 2° and 4°. Participants reported tilt using the left/right arrow key with the dominant hand.

**Timing and blocks.** Each trial began with a fixation period of 500 ms, followed by stimulus presentation for 200 ms. Participants had 1,500 ms to respond and were instructed to maintain fixation throughout. Trial structure was identical in testing and training sessions. Each session comprised 6 blocks of 160 trials (960 trials/day), lasting approximately 1.5 hours. Training days comprised 1 session of 6 blocks each (≈ 15 min per block), totaling 90 minutes per session and about 1.5 hours of training per day.

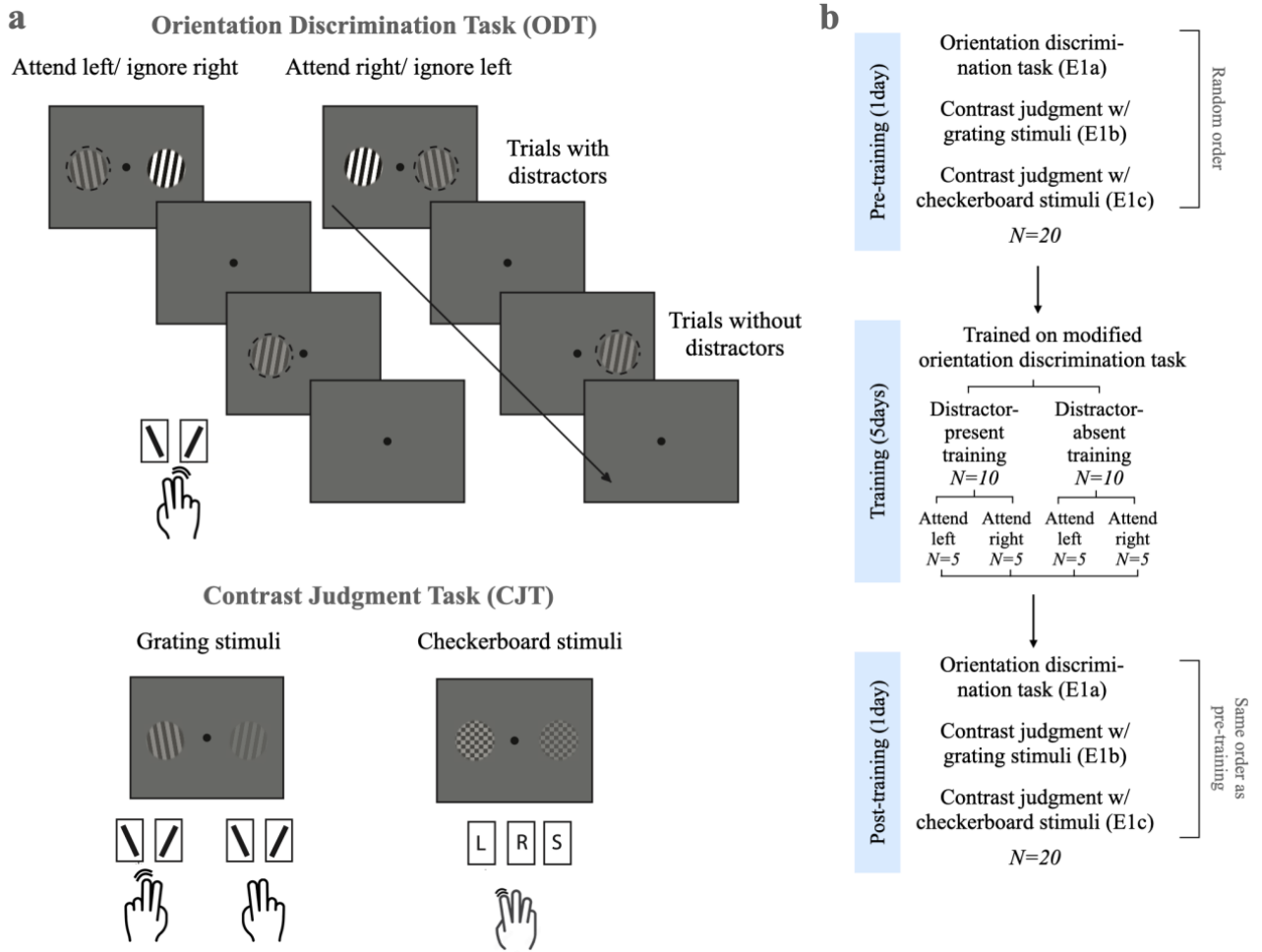

**Figure 1. (a) Tasks.** *Orientation Discrimination Task (ODT):* a low-contrast grating appeared left or right of fixation; on distractor-present trials a high-contrast grating appeared in the opposite hemifield. Participants reported the target's tilt (clockwise/counterclockwise) relative to the vertical meridian. *Contrast Judgment Task (CJT):* two stimuli were shown bilaterally. For gratings, participants used the ipsilateral hand to choose the higher-contrast stimulus and then reported its tilt. For checkerboards, participants indicated which side had higher contrast or that the two were equal. **(b) Seven-day protocol.** Day 1 (pre-training): ODT, CJT-grating, and CJT-checkerboard in randomized order (N = 20). Days 2–6 (training): modified ODT with the target fixed to one side (left or right) and either distractor-present or distractor-absent practice (n = 10 per training condition; n = 5 per trained side). Day 7 (post-training): the same three tasks in the same order as each participant's pre-training session. Data from ODT, CJT-grating, and CJT-checkerboard are reported as Experiments 1a, 1b, and 1c, respectively.

**Data analysis**

**Experiment 1a (ODT).** We used a mixed-measures ANOVA to test the effects of Training Condition (distractor present vs. distractor absent), Target Location (trained vs. untrained), Day (pre vs. post), and Difficulty (easy vs. hard) on ODT performance, with accuracy and response times (RTs) on correct trials as dependent measures. A significant three-way interaction among Training Condition, Target Location, and Day led to follow up analyses. Within each training group, we ran separate repeated-measures ANOVAs on the Target Location by Day interaction. Post-hoc paired t tests then compared change scores (post- minus pre-training) at each target location within each group. We controlled for multiple comparisons using the Holm–Bonferroni method (Holm, 1979). For all experiments, ANOVA effect sizes were reported as generalized eta squared (ges), which represents the proportion of total variance in the dependent variable explained by each factor.

**Experiment 1b (CJT with grating stimuli).** For the CJT with grating stimuli, we computed each participant's point of subjective equality (PSE). For each test contrast level (0.5%, 1.5%, 3%, 4%, 5%, 6%, 8%, 12%, 16%, 24%), we calculated the proportion of trials on which the test was judged higher in contrast than the standard. We then fit a cumulative normal function (mean = 6%, SD = 1%) to obtain the PSE associated with a probability of 0.5 that the test was judged higher than the standard. Only trials with correct tilt responses (clockwise vs. counterclockwise) entered the analysis. Mixed-measures ANOVAs evaluated effects of Training Condition (distractor-present vs. distractor-absent), Day (pre vs. post), and Test Stimulus Location (trained vs. untrained) on PSE values. We ran separate ANOVAs for easy and hard trials. Post-hoc paired t-tests assessed changes in PSE across training conditions and test locations.

**Experiment 1c (CJT with checkerboard stimuli).** For the CJT with checkerboard stimuli, we calculated PSEs using two response types. First, for comparative responses in which participants indicated which stimulus had higher contrast, we estimated the PSE as in Experiment 1b by identifying the test contrast that yielded a probability of 0.5 of being judged higher than the standard. Second, for equality responses where participants judged the two stimuli as equal in contrast, we estimated the PSE as the test contrast with the highest probability of an equality judgment. For each test contrast level (0.5%, 1.5%, 3%, 4%, 5%, 6%, 8%, 12%, 16%, 24%), we computed the proportion of trials judged equal in contrast and fit a normal probability density function with three free parameters (mean = 40%, SD = 1%, α = 0.5) to estimate the PSE. Follow up statistical tests, including mixed-measures ANOVAs and paired t tests, matched the procedures used in Experiment 1b.

**Equipment and settings.** Visual stimuli were presented on a CRT monitor against a gray background 34.51 cd/m2 gamma-corrected with a refresh rate of 120 Hz. The viewing distance was 60 cm in a dim, quiet room. Displays were controlled in MATLAB using Psychtoolbox-3 (Kleiner et al., 2007). Statistical analyses were conducted in MATLAB and R (R Core Team, 2013). All tests were two-tailed with α = .05, and we report 95% CIs throughout.

# Experiment 1 Results

Across seven experimental days, participants achieved 89.84% accuracy (SD = 7.81%) on the orientation discrimination task (ODT) with a mean RT of 535.92 ms (SD = 75.46 ms). No outliers (±3 SD of mean accuracy) were detected, so all data were analyzed. Performance did not differ by the side trained (left vs. right; t(38.68)=0.39, p>0.1), so we collapsed across that factor in subsequent analyses.

## Experiment1a: Orientation Discrimination Task

In Experiment 1a, a mixed-measures ANOVA on accuracy revealed a reliable improvement from pre- to post-training ($F(1,18) = 17.89$, $p < .001$, ges = .13; post = 91.47% vs. pre = 87.12%), indicating that five days of practice enhanced the accuracy of orientation judgments. Accuracy was higher for larger tilt offsets (4° > 2°; $F(1,18) = 202.08$, $p < .001$, ges = .42). Crucially, as illustrated in Figure 2, the improvement in accuracy was larger at the trained than the untrained location (*Day × Location*: $F(1,18) = 5.11$, $p = .03$, ges = .01), and this effect depended on the training regimen (*Day × Location × Training Condition*: $F(1,18) = 5.20$, $p < .05$, ges = .01). Follow-up tests showed that without distractors, accuracy increased broadly across both locations (main effect of Day: $F(1,9) = 9.90$, $p = .01$, ges = .22; no Day × Location interaction, $F(1,9) = 0.00$, $p > .10$; trained: $t(9) = 4.14$, $p < .01$; untrained: $t(9) = 2.34$, $p < .05$). With distractors, improvement was location-specific, emerging at the trained location ($t(9) = 4.03$, $p < .01$) but not the untrained location ($t(9) = 0.65$, $p > .10$). Improved performance at the trained compared to the untrained location is consistent with the significant Day × Location interaction in that group ($F(1,9) = 8.58$, $p = .02$) alongside an overall pre-to-post behavioral gain ($F(1,9) = 8.09$, $p = .02$). Together, these results indicate that incorporating salient distractors during training was necessary to produce spatially selective improvements in behavioral performance. Figure 2 summarizes these accuracy effects, highlighting location-specific improvements after distractor-present training versus bilateral gain after distractor-absent training, with 95% CIs shown.

Correct RTs did not differ between trained and untrained locations ($F(1,18) = 0.503$, $p = .49$), but were faster for easy than hard trials ($F(1,18) = 122.90$, $p < .001$) and faster post- than pre-training ($F(1,18) = 34.21$, $p < .001$; both groups, $t(9) \geq 2.94$, $p \leq .016$). Trials containing distractors were modestly slower by 7.86 ms on average ($F(1,18) = 40.87$, $p < .001$). The three-way interaction on RTs was not significant ($F(1,18) = 1.35$, $p = .26$), indicating that the spatial specificity observed in accuracy did not manifest in changes in RT.

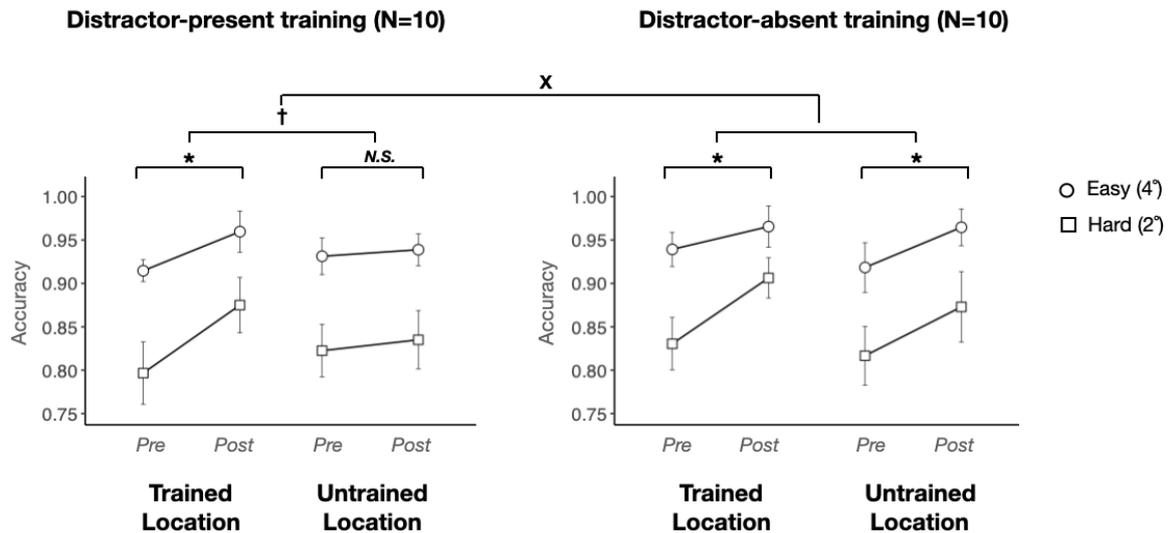

Figure 2. Experiment 1a. Orientation-discrimination accuracy (mean ± 95% CI) from pre- to post-training as a function of Training Condition (distractor-present vs. distractor-absent) and Test Location (trained vs. untrained). Circles = easy (4°); squares = hard (2°). In the distractor-present group (left), accuracy improved at the trained location *but not at the untrained location (N.S.); the within-group Day × Location interaction is significant (†). In the distractor-absent group (right), accuracy improved at both locations*. The across-group Day × Location × Training Condition interaction is significant (X), indicating that training with distractors produced location-specific gains. $N = 10$ per group. * and ** denote significant main effects of day with $p < 0.05$ and $p < 0.01$, respectively.

**Experiment1b: Contrast Judgment with Grating Stimuli**

We examined PSEs in a 2 (Training Condition: distractor-present vs. distractor-absent; between) × 2 (Day: pre vs. post; within) × 2 (Test Location: trained vs. untrained; within) mixed-measures ANOVA. The Training Condition × Day × Test Location interaction was significant, $F(1,18) = 19.95$, $p < .001$, ges = .05, indicating that training altered perceived contrast in a location-dependent manner. After distractor-present training, PSEs decreased at the trained location with participants perceiving the test as higher contrast than the 6% standard (easy: $t(9) = 7.77$, $p < .001$; hard: $t(9) = 7.09$, $p < .001$). In contrast, PSEs increased at the untrained location (easy: $t(9) = 4.06$, $p = .003$; hard: $t(9) = 5.70$, $p < .001$), indicating lower apparent contrast (Fig. 3a). Finally, distractor-absent training produced no reliable PSE changes (easy: $t(9) = 0.14$, $p = .89$; hard: $t(9) = 0.17$, $p = .54$).

Overall, the results showed that distractor-present training produced opposite shifts in perceived contrast, with an increase at the trained location and a decrease at the untrained location. In contrast, distractor-absent training did not produce any significant changes in PSE. This pattern suggests that training with distractors enhances spatial selectivity, sharpening attentional allocation and biasing perceptual appearance of visual stimuli at attended relative to ignored locations (Fig. 3a–c).

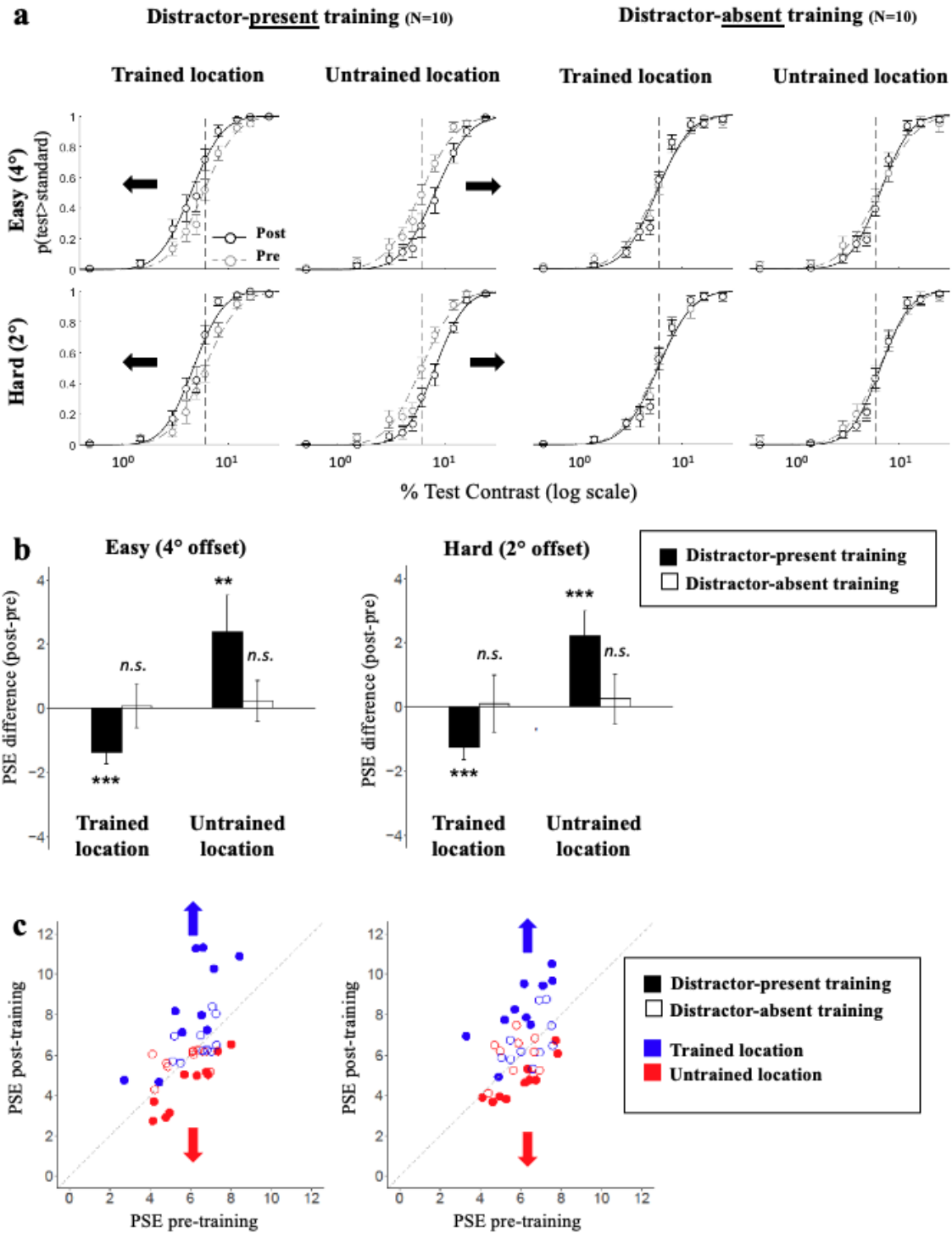

**Figure 3. Experiment 1b (CJT with grating stimuli).** (a) Psychometric functions (mean ± 95% CI) for distractor-present (left) and distractor-absent (right) training, plotted separately for trained and untrained test locations and for easy (4°) and hard (2°) offsets. Open symbols/gray lines = pre; filled symbols/black lines = post. The vertical dashed line marks the 6% standard. Horizontal arrows indicate significant pre-to-post PSE shifts observed only after distractor-present training: leftward at the trained location (higher apparent contrast) and rightward at the untrained location (lower apparent contrast). (b) ΔPSE (post–pre) by group

(solid = distractor-present; open = distractor-absent), difficulty, and location. Negative bars denote lower PSE (higher apparent contrast) and positive bars denote higher PSE (lower apparent contrast). Opponent shifts appear only in the distractor-present group; "n.s." = not significant. Error bars = 95% CI. (c) Individual PSE scatterplots (pre on x, post on y). Filled symbols = distractor-present; open = distractor-absent; blue = trained location, red = untrained location. Points below the diagonal indicate PSE decreases; above indicate increases. After distractor-present training, trained-location points cluster below the diagonal and untrained-location points above, whereas distractor-absent points lie near the diagonal. $N$ = 10 per group. *, **, and *** denote significant main effects of day ($p < 0.05$, $p < 0.01$, and $p < 0.001$, respectively).

**Experiment 1c: Contrast Judgment Task with Checkerboard Stimuli**

We analyzed PSEs for checkerboards in a 2 (Training Condition) × 2 (Day) × 2 (Test Location) design, separating comparative ("which is higher?") and equality ("same contrast?") responses. For comparative responses, the Training Condition × Day × Test Location interaction was significant ($F(1,18) = 21.04$, $p < .001$, ges = .10). After distractor-present training, PSEs decreased at the trained location (leftward shift; $t(9) = 4.42$, $p = .002$), indicating higher apparent contrast, and increased at the untrained location (rightward shift), indicating lower apparent contrast (Fig. 4a). The distractor-absent group showed no reliable change ($t(9) = 0.49$, $p = .63$).

Equality responses showed the same opponent pattern: in the distractor-present group, PSEs shifted leftward at the trained location ($t(9) = 3.87$, $p = .004$) and rightward at the untrained location ($t(9) = 4.29$, $p = .002$), consistent with higher apparent contrast where trained and lower where untrained. No shifts were observed after distractor-absent training (trained: $t(9) = 1.47$, $p = .18$; untrained: $t(9) = 1.59$, $p = .15$).

Thus, even with novel stimuli (i.e., checkerboard instead of grating stimuli) and a response format designed to reduce decisional bias, distractor-present training still produced location-specific changes in perceived contrast, whereas distractor-absent training did not.

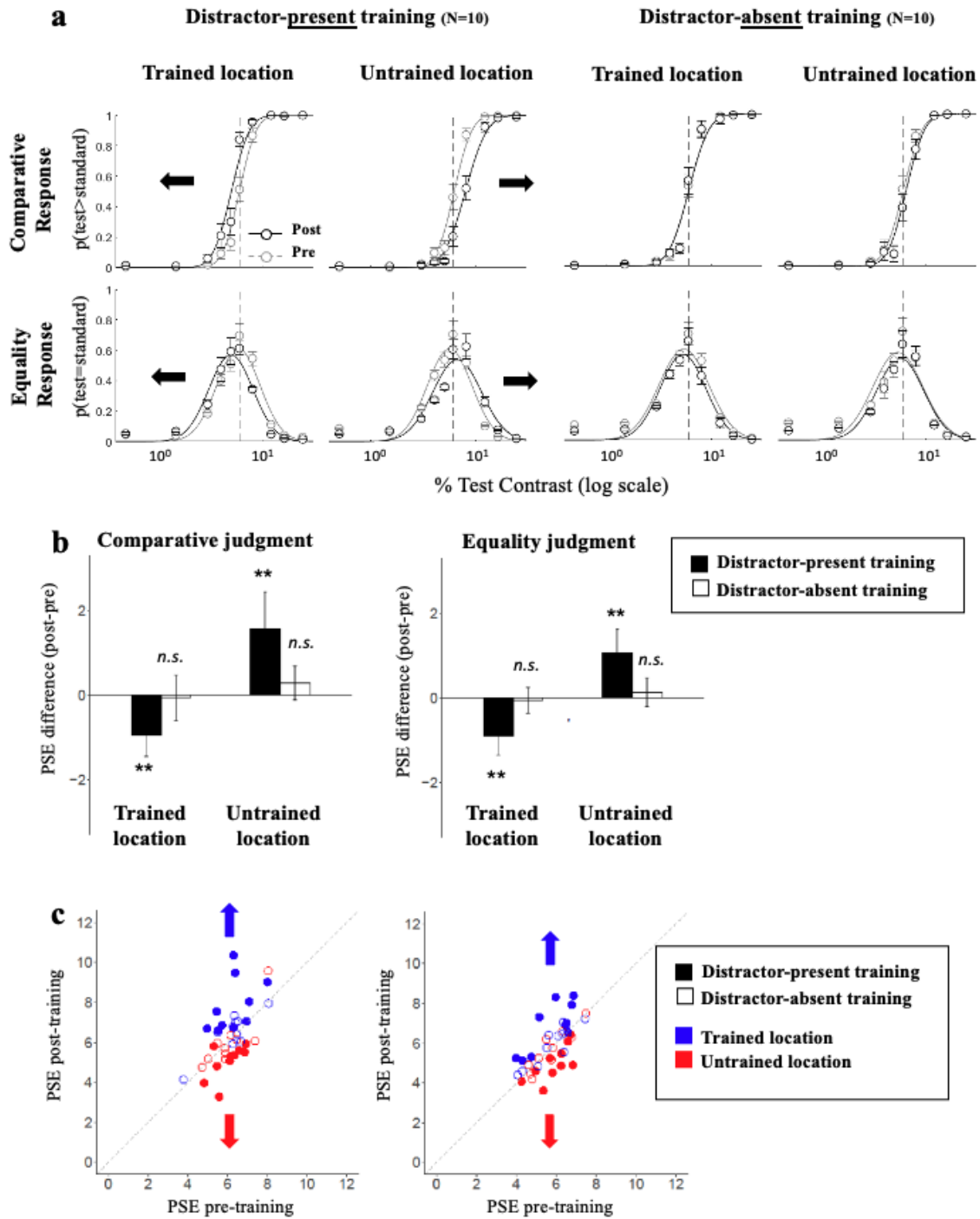

**Figure 4. Experiment 1c (checkerboard CJT).** (a) Psychometric functions (mean ± 95% CI) for distractor-present and distractor-absent training, separated by Test Location (trained vs. untrained) and Response type (top: comparative; bottom: equality). Open symbols/gray curves = pre; filled symbols/black curves = post. The vertical dashed line marks the 6% standard. Horizontal arrows highlight significant pre-to-post PSE shifts that occur only after distractor-present training: leftward at the trained location (higher apparent contrast) and rightward at the untrained location (lower apparent contrast). (b) ΔPSE (post–pre) by group (black = distractor-present; white = distractor-absent), shown separately for comparative and equality judgments and for trained vs. untrained locations. Negative values indicate lower PSE (higher apparent contrast); positive values indicate higher PSE (lower apparent contrast).

Opponent, location-specific shifts are present only in the distractor-present group; "n.s." = not significant. Error bars = 95% CI. (c) Individual PSE scatterplots (pre on x-axis, post on y-axis). Filled symbols = distractor-present; open = distractor-absent; blue = trained location, red = untrained location. Points below the diagonal reflect PSE decreases; above reflect increases. After distractor-present training, trained-location points cluster below and untrained-location points above the diagonal; distractor-absent points lie near the diagonal. $N = 10$ per group. *, **, and *** denote significant main effects of Day ($p < 0.05$, $p < 0.01$, and $p < 0.001$, respectively).

## Experiment 2

The results of Experiment 1 demonstrate that training with distractors produces location-specific changes in appearance that partially generalize to checkerboard stimuli, albeit with reduced magnitude. Experiment 2 was designed to further test the limits of this generalization by examining stimuli that differed from the training target in orientation, contrast, or both. The procedures were identical to Experiment 1, except that on the pre- and post-training days, participants completed three versions of the contrast judgment task (Fig. 5): (i) Orientation Change, in which gratings were tilted relative to the horizontal rather than the vertical meridian; (ii) Contrast Change, in which the standard contrast increased from 6% to 40%; and (iii) Orientation-and-Contrast Change, which combined both manipulations. If distractor-present training induces spatially anchored priority biases rather than purely feature-specific learning, we expected to observe PSE shifts across all variants (specifically, lower PSEs at the trained location and higher at the untrained location). However, based on the partial generalization observed in Experiment 1c, we anticipated that the magnitude of these shifts might decrease as test stimuli diverged more substantially from the training parameters.

**Materials and Method**

**Participants**

Twenty UC San Diego undergraduates (11 female; age = 18-24 years old; all right-handed) took part in Experiment 2. All reported normal or corrected-to-normal vision and provided written informed consent under protocols approved by the UCSD Institutional Review Board.

**Procedure**

Procedures matched Experiment 1 except where noted. Each participant completed seven sessions: a pre-training session (Day 1), five training days (Days 2–6), and a post-training session (Day 7). During training, participants performed the orientation-discrimination task with the target grating consistently presented on either the left or the right side of fixation. The trained side was counterbalanced across participants. Training condition was manipulated between subjects: in the distractor-present group, a high-contrast distractor appeared on the side opposite the target on 50% of trials, whereas in the distractor-absent group no distractors were shown. On Days 1 and 7, participants completed three grating-based contrast-judgment task (CJT) variants in counterbalanced order (Fig. 5). In the Orientation Change variant, gratings were tilted 2° or 4° relative to the horizontal meridian (whereas training used vertical). In the Contrast Change variant, orientation was unchanged from training (vertical), and the standard contrast was increased from 6% to 40% Michelson, and the test contrast was randomly sampled from ten levels (0.5%, 15%, 30%, 35%, 40%, 45%, 55%, 70%, 85%, 100%). In the Orientation + Contrast Change variant, a horizontal stimulus with different contrast levels was used. On every CJT trial participants first used the hand ipsilateral to the chosen side indicated which stimulus had the higher contrast; they then reported whether the higher-contrast grating was tilted clockwise or counterclockwise relative to the meridian relevant for that variant. As in Experiment 1b, the test-stimulus location alternated half-way through each block, and block order was counterbalanced across participants.

**Data analysis**

Data were preprocessed and analyzed exactly as in Experiment 1b. Points of subjective equality (PSEs) were estimated from psychometric fits to the proportion of "test > standard" responses, and mixed-measures ANOVAs assessed effects of Training Condition (between subjects), Day (within subjects), and Test Location (within subjects), with planned paired comparisons probing pre- to post-training changes at each location within each training group.

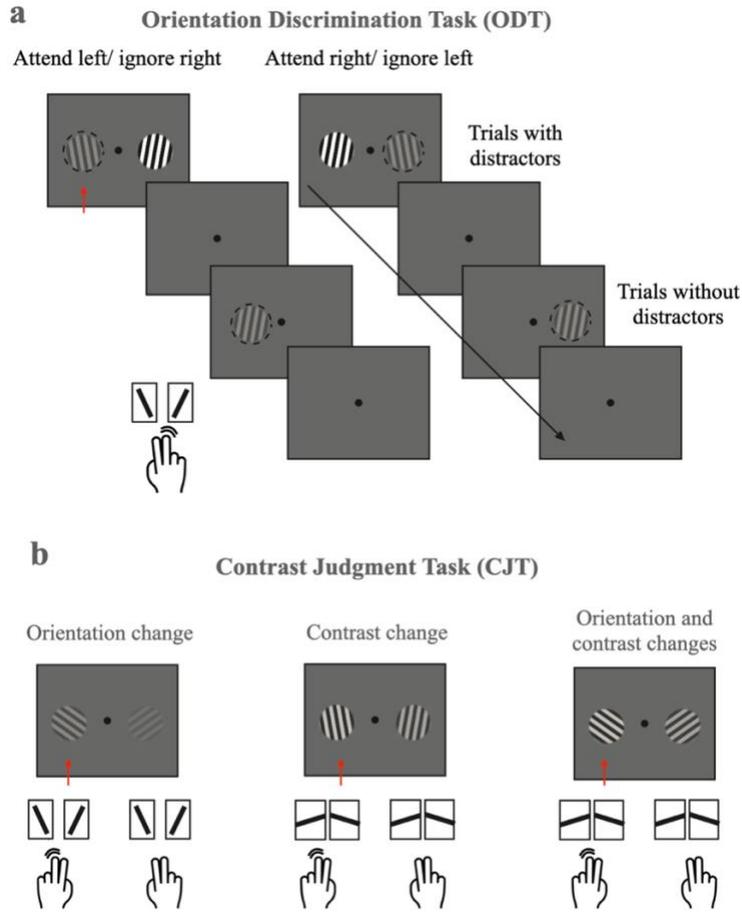

**Figure 5. Experiment 2: tasks and manipulations.** (a) Training (ODT). As in Experiment 1, participants performed the orientation-discrimination task for five days with the target grating consistently on either the left or right of fixation. In the distractor-present condition, a high-contrast distractor appeared on the opposite side on a subset of trials; in the distractor-absent condition no distractor was shown. (b) Pre/Post testing (CJT). On Days 1 and 7, participants completed three grating-based contrast-judgment variants: Orientation Change (tilt defined relative to the horizontal meridian), Contrast Change (standard increased from 6% to 40% with higher test contrasts), and Orientation-and-Contrast Change (both manipulations). Red arrows mark the feature changes relative to the training target. On each trial, participants indicated which side had higher contrast with the ipsilateral hand, then reported the tilt direction (clockwise/counterclockwise) of the higher-contrast grating with the index or middle finger.

## Experiment 2 Results

Across the seven sessions, participants achieved an average ODT accuracy of 91.93% (SD = 3.84%) with a mean RT of 589.32 ms (SD = 106.68 ms). No outliers (±3 SD of mean accuracy) were detected and all data were retained. Performance did not differ by trained side (left vs. right; $t(19.15)=-0.55$, $p>0.1$), so we collapsed across that factor for all analyses.

**Orientation change.** When tilt was defined relative to the horizontal meridian, distractor-present training still produced shifts in perceived contrast (Fig. 6, top; Fig. 7). PSEs decreased at the trained location (easy: $t(9) = 3.00$, $p < .05$; hard: $t(9) = 3.07$, $p < .05$) and increased at the untrained location (easy: $t(9) = 2.64$, $p < .05$; hard: $t(9) = 3.31$, $p = .01$). The distractor-absent group showed no pre–post change (easy: $t(9) = 0.10$, $p = .93$; hard: $t(9) = 1.44$, $p = .18$). Note, however, that the Group × Day interaction on PSE was not reliable for either difficulty (easy: $F(1,9) = 2.91$, $p = .10$; hard: $F(1,9) = 1.41$, $p = .30$). Thus, these between-group effects reflecting the impact of distractors during training on perceived contrast should be interpreted with caution.

**Contrast change.** When the standard increased to 40%, opponent PSE shifts again appeared only after distractor-present training (Fig. 6, middle). There was a leftward shift in PSEs at the trained location (easy: $t(9) = 5.39$, $p < .001$; hard: $t(9) = 6.62$, $p < .001$) and a rightward shift at the untrained location (easy: $t(9) = 4.78$, $p < .01$; hard: $t(9) = 3.98$, $p < .01$). The distractor-absent group showed no change (easy: $t(9) = 0.09$, $p = .90$; hard: $t(9) = -0.04$, $p = 1.00$). Here, the between-subjects Group × Day interaction on PSE was significant (easy: $F(1,9) = 5.55$, $p < .05$; hard: $F(1,9) = 5.78$, $p < .05$), confirming the training-contingent modulation.

**Orientation-and-contrast change.** When both manipulations were applied, the distractor-present group again showed a leftward shift in PSEs at the trained location (easy: $t(9) = 5.08$, $p < .001$; hard: $t(9) = 5.33$, $p < .001$) and a rightward shift at the untrained location (easy: $t(9) = 2.92$, $p < .05$; hard: $t(9) = 2.63$, $p < .05$; Fig. 6, bottom; Fig. 7). No changes emerged after distractor-absent training (trained: easy $t(9) = 0.21$, $p = .80$; hard $t(9) = 0.34$, $p = .70$; untrained: easy $t(9) = 0.35$, $p = .70$; hard $t(9) = 0.60$, $p = .60$). The Group × Day interaction on PSE was significant for easy trials ($F(1,9) = 5.84$, $p < .05$) and marginal for hard trials ($F(1,9) = 3.95$, $p = .08$).

In sum, across three CJT variants that introduced novel feature contexts, distractor-present training reliably yielded opponent, location-specific shifts in perceived contrast. There were lower PSEs at the trained location and higher PSEs at the untrained location, whereas shifts in the distractor-absent training condition were not significant. These effects were strongest in the contrast-change and combined variants, in which reliable Group × Day interactions were observed, and were present but less robust in the orientation-change-only variant, where the between-subject interaction did not reach significance.

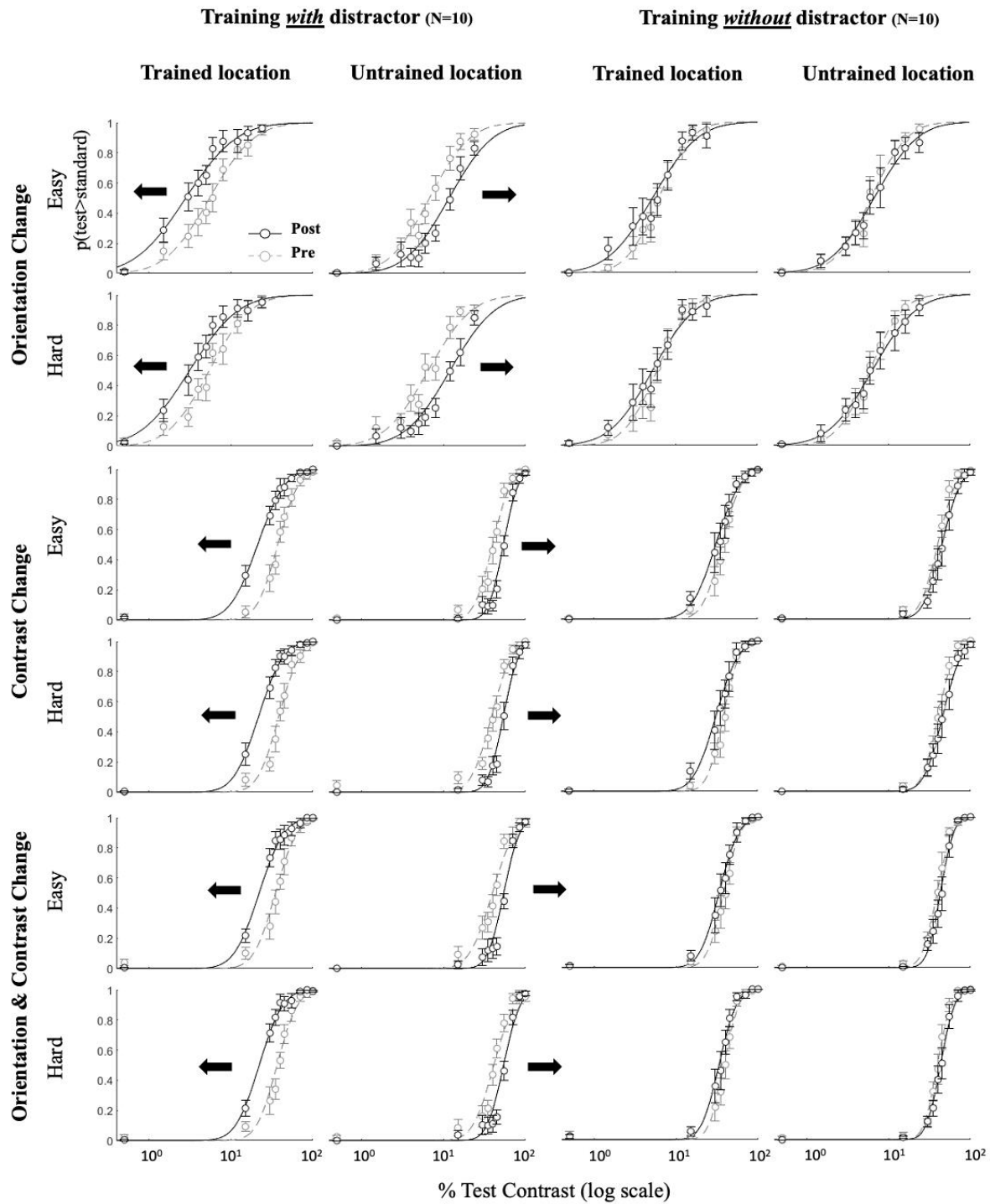

**Figure 6. Experiment 2 (grating CJT with novel-feature variants).** Psychometric functions (mean ± 95% CI) showing the probability that the test appeared higher contrast than the standard as a function of test contrast (log scale). Columns compare training with vs. without a distractor and, within each, trained vs. untrained test locations. Rows show the three task variants (each with easy 4° and hard 2°): Orientation Change, Contrast Change, and Orientation-and-Contrast Change. Open symbols/gray curves = pre; filled symbols/black curves = post. Vertical dashed lines mark the standard contrast (6% for Orientation Change; 40% for Contrast Change and Orientation + Contrast Change). Black arrows highlight reliable pre→post PSE shifts observed only after distractor-present training: leftward at the trained location (higher apparent contrast) and rightward at the untrained location (lower apparent contrast). No systematic shifts are evident after distractor-absent training. *N* = 9 per group. *, **, and *** represent main effects of days with p's <0.05, < 0.01, and <0.001, respectively.

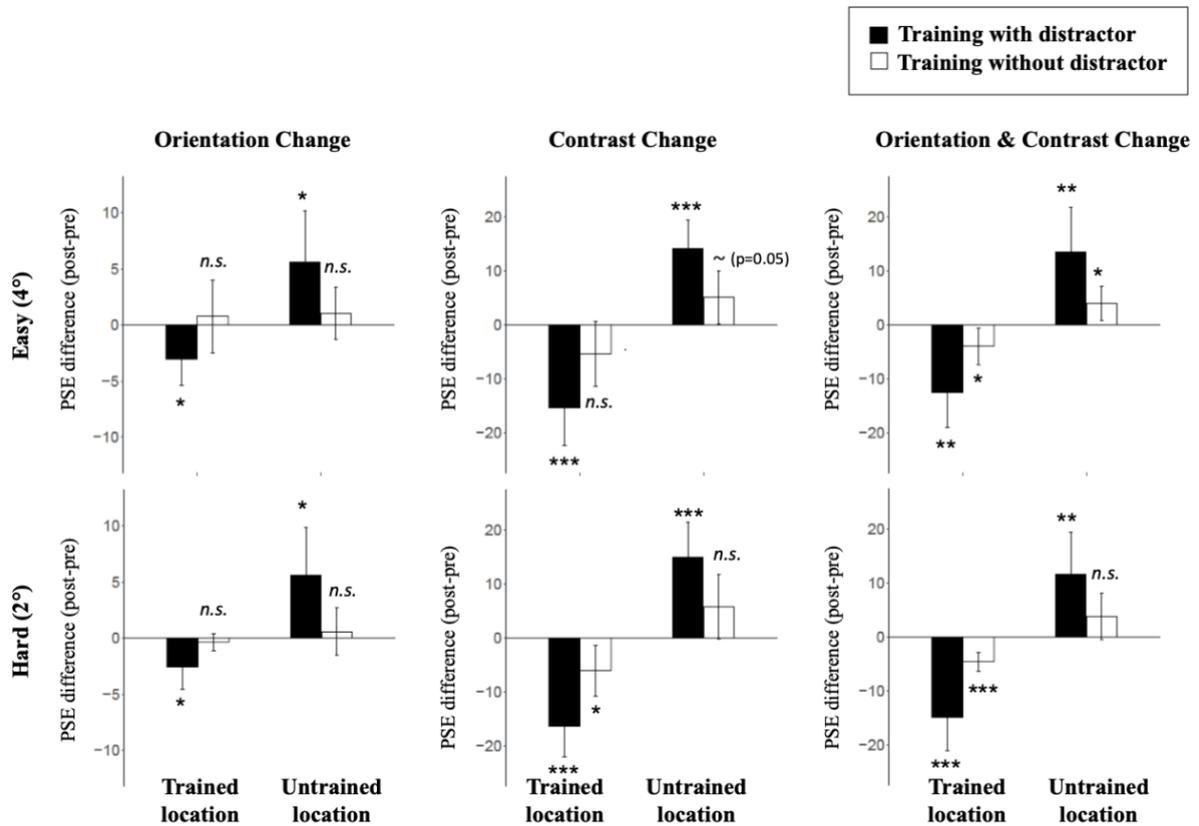

**Figure 7. Experiment 2: ΔPSE (post–pre) by variant, difficulty, and location.** Bar plots show change in PSE for Orientation Change, Contrast Change, and Orientation-and-Contrast Change at Easy (4°) and Hard (2°) offsets, split by Test Location (trained vs. untrained) and Training Group (black = with distractor; white = without). Negative ΔPSE indicates lower PSE (higher apparent contrast); positive ΔPSE indicates higher PSE (lower apparent contrast). After distractor-present training, shifts are opponent and location-specific— decreases at the trained location and increases at the untrained location—whereas distractor-absent training yields values near zero (one marginal case noted, ~$p$ = .05). Asterisks mark within-bar pre-to-post significance (* $p$ < .05; ** $p$ < .01; *** $p$ < .001); "n.s." = not significant. Error bars = 95% CIs. *N* = 9 per group.

## General discussion

A large body of work shows that perceptual learning improves behavioral performance, often in a location- or feature-specific manner (Bao et al., 2010; Byers & Serences, 2014; Itthipuripat et al., 2017; Jehee et al., 2012). Separately, research on selective attention has demonstrated that transient attentional cues can alter subjective appearance on a moment-by-moment basis, with attended stimuli perceived as higher in contrast than physically identical unattended stimuli (Carrasco et al., 2004; Anton-Erxleben & Carrasco, 2013; Liu et al., 2009; Anton-Erxleben et al., 2010; Itthipuripat et al., 2023; Itthipuripat, Vo, et al., 2019). What has remained unclear is whether learning to deploy attention over time, particularly under conditions of sustained competition, can produce enduring changes in how stimuli are experienced. The present findings address this gap by showing that repeatedly resolving attentional competition leaves a lasting, spatially specific imprint on subjective appearance.

Across two experiments, attention training under sensory competition produced a reliable opponent shift in perceived contrast. Stimuli presented at the trained location were subsequently perceived as higher in contrast, whereas stimuli presented at the untrained location were perceived as lower. Crucially, this push–pull pattern emerged only when training required participants to ignore a salient distractor. Training without competition improved discrimination accuracy but did not alter appearance. Moreover, these appearance shifts generalized beyond the trained stimulus to novel judgment formats and stimulus contexts, although with reduced magnitude as similarity to the training context decreased. Together, these findings indicate that subjective visual experience is not only momentarily shaped by attentional state but can be durably recalibrated by the history of resolving competition between relevant and irrelevant information.

This pattern of results is consistent with biased-competition accounts of attention, which propose that selection involves a redistribution of perceptual influence across competing representations rather than uniform enhancement (Desimone & Duncan, 1995; Bisley & Goldberg, 2010; Gottlieb & Balan, 2010; Jerde & Curtis, 2013; Sprague et al., 2015; Sprague & Serences, 2013). From a psychological perspective, repeated exposure to attentional conflict may bias spatial priority signals such that information from consistently selected locations contributes more strongly to appearance judgments, while information from consistently ignored locations contributes less. This opponent structure provides a parsimonious explanation for why the trained location showed enhanced appearance while the untrained location showed a corresponding reduction. Importantly, this pattern distinguishes the present effects from generic practice-related improvements and highlights the critical role of competition, rather than repetition alone, in reshaping appearance.

The present findings also extend prior work demonstrating that attention can alter appearance. Earlier studies relied primarily on transient cues that bias perception within individual trials (Carrasco et al., 2004; Anton-Erxleben et al., 2010; Anton-Erxleben & Carrasco, 2013). Here, we show that comparable biases in appearance can emerge through learning, persist beyond the training phase, and be expressed even in the absence of attentional cues. The use of equality judgments and non-grating stimuli, both designed to reduce reliance on fixed response strategies, further supports the conclusion that the observed shifts in point of subjective equality reflect changes in subjective appearance rather than a unitary decision criterion (Anton-Erxleben et al., 2010; Itthipuripat et al., 2023). Critically, the opponent direction of shifts across locations is difficult to explain by a single global criterion shift. Thus, attention's influence on appearance is not limited to transient attentional states but can become an enduring property of perceptual experience shaped by prior selection history.

These findings also connect naturally to the perceptual learning literature. Perceptual learning is often spatially specific but can generalize under conditions that engage attention or increase representational overlap between training and test (Bao et al., 2010; Wang et al., 2012; Xiao et al., 2008). Prominent theories propose that such learning reflects selective reweighting of sensory signals guided by task demands and top-down control (Ahissar & Hochstein, 2004; Dosher & Lu, 1999, 2017; Petrov et al., 2005). In the present study, training without distractors produced broad performance gains, whereas training under competition yielded spatially specific improvements accompanied by opponent shifts in appearance. This dissociation suggests that attention gates plasticity differently depending on task context, with sustained competition biasing how spatial information is weighted in appearance judgments rather than simply improving overall sensitivity (Byers & Serences, 2014; Itthipuripat et al., 2017; Itthipuripat & Serences, 2016).

At a mechanistic level, the results are compatible with normalization-based accounts of attentional modulation, in which enhancement at one location is coupled with reduced influence from competing locations (Reynolds et al., 1999; Martínez-Trujillo & Treue, 2002; Itthipuripat, Garcia, et al., 2014). Repeated deployment of attention under competition may bias this balance over time, producing lasting changes in the relative contribution of information from different spatial locations. The observation that appearance shifts were strongest when test stimuli shared key dimensions with the training context, and weaker when they diverged, suggests that training recalibrates spatial priority signals that modulate population-level responses without fully overriding feature-selective sensory representations (Sprague et al., 2015; Sprague & Serences, 2013).

Several limitations warrant consideration. The present conclusions are based on behavioral measures of performance and appearance rather than direct neural recordings, and thus cannot localize the representational locus of the observed changes. Although equality judgments reduce decisional bias, formal signal-detection or computational modeling could further dissociate changes in sensitivity-like components from location-dependent priors or criteria in appearance judgments. Future work combining this paradigm with neuroimaging or electrophysiological measures, along with model-based analyses of contrast-response functions and spatial population codes, would enable stronger inferences about the neural mechanisms underlying learned changes in appearance (Byers & Serences, 2014; Foster et al., 2021; Itthipuripat & Serences, 2016; Pestilli et al., 2011; Sprague & Serences, 2013).

## Conclusion

In conclusion, training selective attention under competition can durably reshape subjective visual experience. Only training that required participants to repeatedly resolve sensory competition produced opponent shifts in perceived contrast, with enhancement at the trained location and reduced appearance at the untrained location. In contrast, training without competition improved discrimination accuracy without altering appearance. These learned biases generalized across judgment formats and stimulus classes, although with reduced magnitude as test stimuli diverged from the training context. Together, these results provide direct evidence that the history of resolving attentional competition can recalibrate perceptual weighting, altering not only how efficiently information is processed but how the visual world is experienced.

## Declarations

**Funding:** This work was funded by NIH R01-EY025872 and a James S. McDonnell Foundation award to John T. Serences. This project was also funded by the National Research Council of Thailand (NRCT) grant (years 2021-2025), the NRCT grant for Mid-career Talented Researchers (year 2025), the Thailand Science Research and Innovation (TSRI) Fundamental Fund (years 2020-2025), the Project Management Units (years 2022-2024), the Asahi Glass Foundation grant, the KMUTT Partnering initiative grant, the startup fund for junior researchers at King Mongkut's University of Technology Thonburi (KMUTT), and the KMUTT's Frontier Research Unit Grant for Neuroscience Center for Research and Innovation to Sirawaj Itthipuripat. The project was also supported by the NRCT grant for young talented researchers (years 2024-2025), TSRI Fundamental Fund (years 2024-2025), and the International Brain Research Organization (IBRO) Rising Stars Award (years 2023-2025) to Thitaporn Chaisilprungraung.

**Conflicts of interest:** The authors have no competing interests to declare that are relevant to the content of this article.

**Ethics approval:** All procedures were performed in accordance with the ethical standards of the Declaration of Helsinki. The study was approved by the Institutional Review Board of the University of California, San Diego.

**Consent to participate:** Informed consent was obtained from all individual participants included in the study.

**Consent for publication:** Not applicable.

**Availability of data and materials:** The datasets generated during and/or analysed during the current study are available from the corresponding author on reasonable request.

**Code availability:** The analysis code supporting the findings of this study is available from the corresponding author on reasonable request.

**Authors' contributions:** T.C. performed data analysis, prepared figures, drafted the manuscript, and led all subsequent revisions. P.S. contributed to data analysis. J.T.S. supervised the project and co-wrote/edited the manuscript. S.I. conceived the project, collected data, supervised the research, contributed to data analysis, and co-wrote/edited the manuscript. We thank Kai-Yu Chang for her help with data collection.

# References

Ahissar, M., & Hochstein, S. (2004). The reverse hierarchy theory of visual perceptual learning. *Trends in Cognitive Sciences, 8*(10), 457–464. https://doi.org/10.1016/j.tics.2004.08.011

Anton-Erxleben, K., Abrams, J., & Carrasco, M. (2010). Evaluating comparative and equality judgments in contrast perception: Attention alters appearance. *Journal of Vision, 10*(11), 6. https://doi.org/10.1167/10.11.6

Anton-Erxleben, K., & Carrasco, M. (2013). Attentional enhancement of spatial resolution: Linking behavioural and neurophysiological evidence. *Nature Reviews Neuroscience, 14*(3), 188–200. https://doi.org/10.1038/nrn3443

Bao, M., Yang, L., Rios, C., He, B., & Engel, S. A. (2010). Perceptual learning increases the strength of the earliest signals in visual cortex. *Journal of Neuroscience, 30*(45), 15080–15084. https://doi.org/10.1523/JNEUROSCI.5703-09.2010

Bisley, J. W., & Goldberg, M. E. (2010). Attention, intention, and priority in the parietal lobe. *Annual Review of Neuroscience, 33*, 1–21. https://doi.org/10.1146/annurev-neuro-060909-152823

Byers, A., & Serences, J. T. (2014). Enhanced attentional gain as a mechanism for generalized perceptual learning in human visual cortex. *Journal of Neurophysiology, 112*(5), 1217–1227. https://doi.org/10.1152/jn.00353.2014

Carrasco, M., Ling, S., & Read, S. (2004). Attention alters appearance. *Nature Neuroscience, 7*(3), 308–313. https://doi.org/10.1038/nn1194

Desimone, R., & Duncan, J. (1995). Neural mechanisms of selective visual attention. *Annual Review of Neuroscience, 18*, 193–222. https://doi.org/10.1146/annurev.ne.18.030195.001205

Dosher, B. A., & Lu, Z.-L. (1999). Mechanisms of perceptual learning. *Vision Research, 39*(19), 3197–3221. https://doi.org/10.1016/S0042-6989(99)00059-0

Dosher, B. A., & Lu, Z.-L. (2017). Visual perceptual learning and models. *Annual Review of Vision Science, 3*, 343–363. https://doi.org/10.1146/annurev-vision-102016-061249

Foster, J. J., Thyer, W., Wennberg, J. W., & Awh, E. (2021). Covert attention increases the gain of stimulus-evoked population codes. *Journal of Neuroscience, 41*(8), 1802–1815. https://doi.org/10.1523/JNEUROSCI.1499-20.2020

Gobell, J., & Carrasco, M. (2005). *Attention alters the appearance of spatial frequency and gap size*. Psychological Science, 16(8), 644–651. https://doi.org/10.1111/j.1467-9280.2005.01588.x

Gottlieb, J., & Balan, P. (2010). Attention as a decision in information space. *Trends in Cognitive Sciences, 14*(6), 240–248. https://doi.org/10.1016/j.tics.2010.03.001

Itthipuripat, S., Cha, K., Byers, A., & Serences, J. T. (2017). Two different mechanisms support selective attention at different phases of training. *PLOS Biology, 15*(6), e2001724. https://doi.org/10.1371/journal.pbio.2001724

Itthipuripat, S., Garcia, J. O., Rungratsameetaweemana, N., Sprague, T. C., & Serences, J. T. (2014). Changing the spatial scope of attention alters patterns of neural gain in human cortex. *Journal of Neuroscience, 34*(1), 112–123. https://doi.org/10.1523/JNEUROSCI.3943-13.2014

Itthipuripat, S., Phangwiwat, T., Wiwatphonthana, P., Sawetsuttipan, P., Chang, K.-Y., Störmer, V. S., Woodman, G. F., & Serences, J. T. (2023). Dissociable neural mechanisms underlie the effects of attention on visual appearance and response bias. *Journal of Neuroscience, 43*(39), 6628–6652. https://doi.org/10.1523/JNEUROSCI.0266-23.2023

Itthipuripat, S., & Serences, J. T. (2016). Integrating levels of analysis in systems and cognitive neurosciences: Selective attention as a case study. *The Neuroscientist, 22*(3), 225–237. https://doi.org/10.1177/1073858415603312

Itthipuripat, S., Vo, V. A., Sprague, T. C., & Serences, J. T. (2019). Value-driven attentional capture enhances distractor representations in early visual cortex. *PLOS Biology, 17*(8), e3000186. https://doi.org/10.1371/journal.pbio.3000186

Jehee, J. F. M., Ling, S., Swisher, J. D., van Bergen, R. S., & Tong, F. (2012). Perceptual learning selectively refines orientation representations in early visual cortex. *Journal of Neuroscience, 32*(47), 16747–16753. https://doi.org/10.1523/JNEUROSCI.6112-11.2012

Jerde, T. A., & Curtis, C. E. (2013). Maps of space in human frontoparietal cortex. *Journal of Physiology-Paris, 107*(6), 510–516. https://doi.org/10.1016/j.jphysparis.2013.04.002

Liu, T., Abrams, J., & Carrasco, M. (2009). Voluntary attention enhances contrast appearance. *Psychological Science, 20*(3), 354–362. https://doi.org/10.1111/j.1467-9280.2009.02300.x

Martínez-Trujillo, J. C., & Treue, S. (2002). Attentional modulation strength in cortical area MT depends on stimulus contrast. *Neuron, 35*(2), 365–370. https://doi.org/10.1016/S0896-6273(02)00778-X

Pestilli, F., Carrasco, M., Heeger, D. J., & Gardner, J. L. (2011). Attentional enhancement via selection and pooling of early sensory responses in human visual cortex. *Neuron, 72*(5), 832–846. https://doi.org/10.1016/j.neuron.2011.09.025

Petrov, A. A., Dosher, B. A., & Lu, Z.-L. (2005). The dynamics of perceptual learning: An incremental reweighting model. *Psychological Review, 112*(4), 715–743. https://doi.org/10.1037/0033-295X.112.4.715

Reynolds, J. H., Chelazzi, L., & Desimone, R. (1999). Competitive mechanisms subserve attention in macaque areas V2 and V4. *Journal of Neuroscience, 19*(5), 1736–1753. https://doi.org/10.1523/JNEUROSCI.19-05-01736.1999

Sagi, D. (2011). *Perceptual learning in vision research*. Vision Research, 51(13), 1552–1566. https://doi.org/10.1016/j.visres.2010.10.019

Sprague, T. C., Saproo, S., & Serences, J. T. (2015). Visual attention mitigates information loss in small- and large-scale neural codes. *Trends in Cognitive Sciences, 19*(4), 215–226. https://doi.org/10.1016/j.tics.2015.02.005

Sprague, T. C., & Serences, J. T. (2013). Attention modulates spatial priority maps in the human occipital, parietal and frontal cortices. *Nature Neuroscience, 16*(12), 1879–1887. https://doi.org/10.1038/nn.3574

Wang, R., Zhang, J.-Y., Klein, S. A., Levi, D. M., & Yu, C. (2012). Task relevancy and demand modulate double-training enabled transfer of perceptual learning. *Vision Research, 61*, 33–38. https://doi.org/10.1016/j.visres.2012.03.018

Watanabe, T., & Sasaki, Y. (2015). *Perceptual learning: Toward a comprehensive theory*. Annual Review of Psychology, 66, 197–221. https://doi.org/10.1146/annurev-psych-010814-015214

Xiao, L.-Q., Zhang, J.-Y., Wang, R., Klein, S. A., Levi, D. M., & Yu, C. (2008). Complete transfer of perceptual learning across retinal locations enabled by double training. *Current Biology, 18*(24), 1922–1926. https://doi.org/10.1016/j.cub.2008.10.030